\documentclass[11pt]{article}
\usepackage[margin=1in]{geometry}
\usepackage{amsmath,amssymb}
\usepackage{graphicx}
\usepackage{booktabs}
\usepackage{siunitx}
\usepackage{authblk}
\usepackage[hidelinks]{hyperref}
\usepackage{caption}
\usepackage{float}
\usepackage{lineno}
\usepackage{setspace}

\newcommand{\degC}{\textsuperscript{o}C}

\title{Network amplification of dengue declines as endemicity rises: climate-adjusted directional spread across Costa Rican cantons, 1993--2012}

\author[1]{Fabio Sanchez\thanks{Corresponding author: fabio.sanchez@ucr.ac.cr}}
\affil[1]{Centro de Investigaci\'on en Matem\'atica Pura y Aplicada (CIMPA) and Escuela de Matem\'atica, Universidad de Costa Rica, San Jos\'e, Costa Rica}

\date{\today}

\begin{document}
\maketitle
\onehalfspacing

\begin{abstract}
\noindent
\textbf{Background:} In Costa Rica, dengue is reported and controlled at the canton level, and outbreaks in one canton are often followed by outbreaks in others. Climate models describe where conditions favor transmission but not how dengue moves \emph{between} places, the directional, between-place spread that shapes where an outbreak travels next.

\noindent
\textbf{Methods:} From weekly case counts for all 81 cantons (1993--2012; \num{246524} cases) we reconstructed a canton-to-canton spread map using the roughly three-week dengue generation interval, removed the shared seasonal and climatic signal so that only direction-specific spread remained, and summarized it by the receiving and source cantons, an amplification factor, and a directionality index, tracked over five-year windows.

\noindent
\textbf{Results:} Climate-adjusted spread is strongly directional and concentrates in the lowland Caribbean and Pacific cantons (Lim\'on, Matina, Gu\'acimo, Garabito, Orotina). A local outbreak is amplified about three- to fourfold across the network even though overall transmission is not growing. This amplification was greatest during the emergence phase of the 1990s and declined markedly as annual reported cases increased, while the \emph{direction} of spread remained fixed; the decline persists after controlling for the broadening of surveillance coverage.

\noindent
\textbf{Conclusions:} Routine surveillance alone can map which cantons tend to receive dengue and the pathways along which it appears to spread, providing one potential input for prioritizing surveillance and vector control, particularly when a serotype or the disease itself is newly establishing. As a historical description of average behavior over multi-year windows, it is a planning input whose prospective value remains to be tested.
\end{abstract}



\section{Introduction}

Dengue is the most important mosquito-borne viral disease of the Americas \cite{Bhatt2013}, and both the range of its mosquito vectors \cite{Kraemer2015} and its burden in Central America are expected to grow under continued warming \cite{Colon2018,Childs2024}. Because temperature, humidity, and rainfall shape the biology of the \emph{Aedes} vector, dengue is strongly climate-sensitive, and recent work has produced increasingly fine-grained descriptions of the climate--dengue relationship for Costa Rica, including district-level projections of how climatic suitability for transmission may shift by mid-century \cite{Barboza2023,Hidalgo2026}. These climate-based assessments answer a crucial question, \emph{where} conditions will favor dengue, and are essential for long-range adaptation planning.

They do not, however, describe how dengue moves \emph{between} places in the short term. Outbreaks are seeded by the introduction of infected people and, once established in one location, spread to others through human movement and connectivity rather than through local climate alone \cite{Grenfell2001,Stoddard2013,Wesolowski2015}. This between-place spread is a principal driver of where an emerging outbreak will go next, and it is the dimension along which surveillance and pre-emptive vector control can be targeted before cases appear. Two features of Costa Rica make this tractable. First, dengue is a notifiable disease, and case reports are compiled at the canton level, the administrative unit where control operations are organized. Second, the country experienced a well-defined natural experiment: dengue was reintroduced in 1993 into a population with little recent exposure \cite{PAHO1994,Troyo2006}, and the following two decades of weekly, canton-level surveillance capture the entire arc from emergence to endemicity.

Here we use that record to build a data-driven map of directional dengue spread between cantons. Our aims are threefold: (1) to reconstruct, from routine surveillance alone, which cantons act as sources and which as receivers of dengue; (2) to quantify how strongly a local outbreak is amplified as it propagates through the canton network; and (3) to determine how this between-canton spread changed as dengue became established. Throughout, we take deliberate care to separate the direction-specific, place-to-place signal from the artificial synchrony induced by shared climate, so that what we report is the between-canton association that remains after removing shared climatic co-movement, which we interpret as evidence consistent with directional spread.

\section{Methods}

\subsection{Dengue surveillance data}

We used weekly counts of reported dengue cases for the 81 cantons of Costa Rica compiled by the Ministry of Health from 1993 through mid-2012, comprising $T=\num{1018}$ epidemiological weeks and $\num{246524}$ reported cases across cantons, together with annual cantonal population estimates. The series begins at the 1993 reintroduction, so the earliest weeks reflect an emerging disease in a largely susceptible population. For regional summaries, we grouped cantons into Costa Rica's seven provinces (San Jos\'e, Alajuela, Cartago, Heredia, Guanacaste, Puntarenas, and Lim\'on) using the standard canton coding.

\subsection{Reconstructing canton-to-canton spread}

Dengue transmission runs through a human--mosquito--human cycle, so cases in one location are followed, after a delay, by new cases they generate. The delay, the generation interval, is centered near three weeks for dengue when the human infectious period and the mosquito's incubation and biting are combined \cite{Chan2012}. We used this known timing to reconstruct spread, drawing on established approaches that infer transmission from case timing and a known generation interval \cite{Wallinga2004}. For each canton we measured how strongly its weekly case count is predicted by the recent, generation-interval-weighted case history of every canton, including itself. Writing $I_i(t)$ for cases in canton $i$ at week $t$, and $\Lambda_j(t)=\sum_{s} w_s\,I_j(t-s)$ for the recent transmission pressure originating in canton $j$ (with $w_s$ the weekly generation-interval weights), the reconstruction is
\begin{equation}
  I_i(t)\ \approx\ \sum_{j=1}^{81} C_{ij}\,\Lambda_j(t),
  \label{eq:C}
\end{equation}
where the entry $C_{ij}$ measures how strongly cases in canton $j$ are followed by cases in canton $i$. The generation-interval weights $w_s$ were the discretized gamma density with mean 19 days and standard deviation 6 days, evaluated over $S=8$ weekly lags, and each row of $C$ was fit on the usable weeks $t>S$.

The matrix $C$ is a data-driven map of canton-to-canton spread, estimated one row at a time by regressing a canton's weekly incidence on the generation-interval-weighted histories of all 81 cantons. Because case pressures and spread strengths are non-negative, we constrained $C_{ij}\ge 0$ and added a Tikhonov (ridge) penalty $\lambda\sum_j C_{ij}^2$ to stabilize the inversion, which is otherwise ill-conditioned because neighboring cantons have highly correlated histories. The constrained, penalized problem was solved as an augmented non-negative least-squares system, stacking $\sqrt{\lambda}\,\mathbf{I}$ beneath the design matrix of lagged pressures and zeros beneath the response. Rather than tune $\lambda$ separately for each canton by cross-validation, we fixed it at a single scale-adaptive value, $\lambda=\alpha\,\operatorname{tr}(G^\top G)/81$ with $\alpha=0.05$, that is, five percent of the mean squared magnitude of the lagged-pressure predictors ($G$ denoting the matrix whose columns are the $\Lambda_j(t)$). A common $\lambda$ keeps the penalty comparable across cantons and, in the temporal analysis, across time windows; we examine the influence of this choice below. The matrix diagnostics that summarize the operator follow a companion methodological study, currently in preparation \cite{SanchezRzero}.

\subsection{Separating true spread from shared-climate synchrony}

Cantons rise and fall partly in unison simply because they share weather: a wet, warm season lifts transmission across much of the country at once. This shared forcing can appear, spuriously, as one canton ``spreading'' dengue to another when in fact both are responding to the same climate \cite{Bjornstad1999}. To isolate the direction-specific signal we worked on log-transformed incidence, $\log(1+\text{cases})$, and removed the shared signal in two steps. First, we subtracted each canton's week-of-year climatology, the mean log-incidence in each of the 52 calendar weeks computed over the whole record, which removes the seasonal cycle that cantons at similar latitude and elevation share. Second, from the deseasonalized residuals we removed the $k$ leading modes of nationwide co-movement, that is, the $k$ leading left singular vectors from a single singular value decomposition of the deseasonalized matrix, removed together as one rank-$k$ projection (the decomposition is computed once and not recomputed between removals); these modes capture the year-to-year climatic forcing that raises or lowers transmission across many cantons at once. We used $k=2$. On the resulting anomalies we then re-estimated the coupling operator by the same generation-interval regression as in Eq.~\eqref{eq:C}; because the anomalies are signed, this climate-adjusted operator $A$ was fit by ordinary Tikhonov ridge (the non-negativity constraint applies only to the count-scale map $C$), using the same penalty $\lambda=\alpha\,\operatorname{tr}(G^\top G)/81$, $\alpha=0.05$. For the static, full-record operator both removals were applied to the entire series; for the temporal analysis the seasonal climatology was estimated once on the full record while the $k$ common modes were removed \emph{within} each five-year window, so that the operator in each window is net of the co-movement present in that window. By construction, the residual captures the direction-specific association that is \emph{not} explained by shared climate, precisely the component that climate suitability models omit. We interpret this residual as consistent with connectivity-driven spread, while noting (Section~\ref{sec:limits}) that it remains an association and may retain contributions from drivers correlated with, but not reducible to, shared weather. The conclusions are stable in the number of modes removed: for $k\in\{1,2,3\}$ the windowed amplification factor spanned roughly $2.1$ to $3.7$ and the identity of the top receiving cantons was preserved, with the receiver ranking correlated at Spearman $\rho\approx0.76$ to $0.85$ across the three choices.

\subsection{Summaries and descriptive analyses}

From the climate-adjusted operator $A$ we computed three quantities. \emph{Receiving and source cantons}: receivers and sources are identified by the magnitudes of the dominant right and left eigenvectors of $A$, respectively, each normalized to sum to one; the net difference between a canton's outgoing and incoming off-diagonal strength, $\sum_{i}|A_{ij}|-\sum_{i}|A_{ji}|$, summarizes its role. \emph{Amplification factor}: the largest singular value (spectral norm) of $A$, the greatest factor by which the network can magnify a spatial perturbation over one generation interval; a value above one means the network amplifies a local introduction even when the operator's dominant eigenvalue, a network-wide reproduction proxy in the sense of the next-generation matrix \cite{Diekmann1990,vandenDriessche2002}, is below one. This transient amplification is the network analog of reactivity in dynamical systems \cite{NeubertCaswell1997}. It is computed on the full $81\times81$ operator, including the diagonal, as the reactivity definition requires, rather than on the off-diagonal part alone; the two are close in practice,\footnote{On the full record, the spectral norm is $1.78$ for the full operator versus $1.63$ with the diagonal removed, both well above the dominant eigenvalue of $0.90$.} confirming that the amplification is carried by between-canton coupling and is not an artifact of within-canton persistence. We report windowed estimates, since the operator is re-estimated within each five-year window; pooling the full record yields a smaller value, because the longer series permits more complete removal of the shared signal. The amplification magnitude scales smoothly with the ridge penalty (windowed values of about $3$ to $5$ at $\alpha=0.02$, $2.4$ to $3.6$ at $\alpha=0.05$, and $1.8$ to $2.7$ at $\alpha=0.1$) but stays above one throughout, and the receiver ranking is insensitive to the penalty over this range (Spearman $\rho\ge0.92$ against the $\alpha=0.05$ estimate); we report $\alpha=0.05$. The temporal decline is likewise insensitive to the penalty: for $\alpha$ from $0.02$ to $0.10$ the windowed amplification falls monotonically with time (correlation $\approx-0.95$) by a comparable relative amount ($32$ to $39\%$). \emph{Directionality index}: the fraction of the operator that is one-way rather than reciprocal, computed as $\lVert(A-A^\top)/2\rVert_F/\lVert A\rVert_F$, the Frobenius norm of the antisymmetric part relative to the whole. We estimated all three on the full record and on five-year sliding windows (stepped annually) to track change over time, associating each window with its mean annual case burden. As descriptive context, we also examined the national epidemic curve; the full space--time structure of incidence; seasonal cycles and total burden by province; and the sequence in which dengue established across cantons (defined as the week in which a canton first accumulated at least five reported cases).

\section{Results}

\subsection{The epidemic record and the spatial invasion}

\begin{figure}[H]
  \centering
  \includegraphics[width=\textwidth]{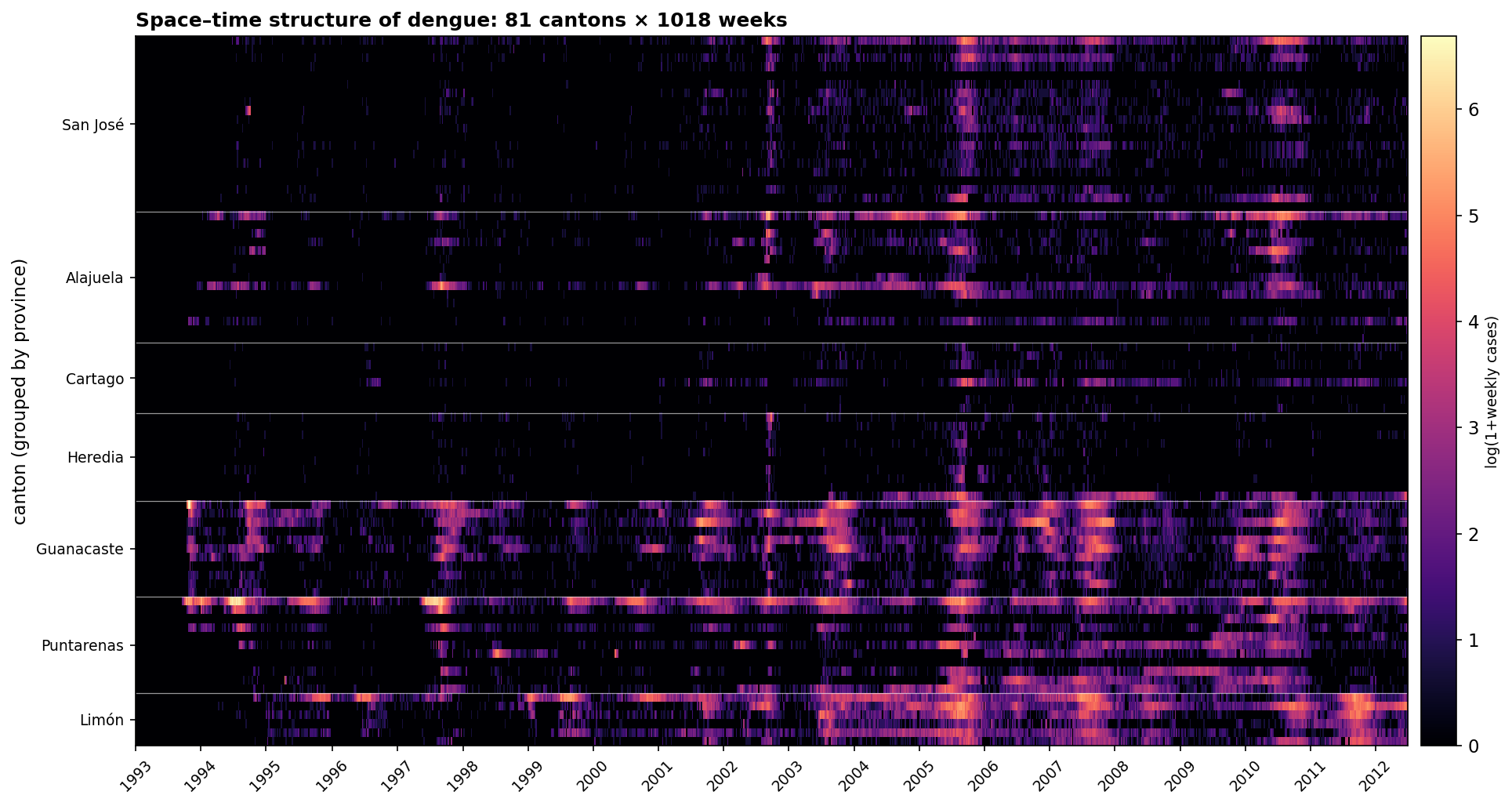}
  \caption{Space--time structure of dengue: weekly incidence for all 81 cantons, grouped by province. Color encodes $\log(1+\text{weekly reported cases})$, with darker shades higher; the logarithmic scale compresses the wide dynamic range so that low-level circulation stays visible alongside major epidemics. Dengue first establishes itself along the Pacific (Guanacaste, Puntarenas) and Caribbean (Lim\'on) lowlands; the Central Valley provinces intensify only from the mid-2000s onward.}
  \label{fig:spacetime}
\end{figure}

\begin{figure}[H]
  \centering
  \includegraphics[width=0.72\textwidth]{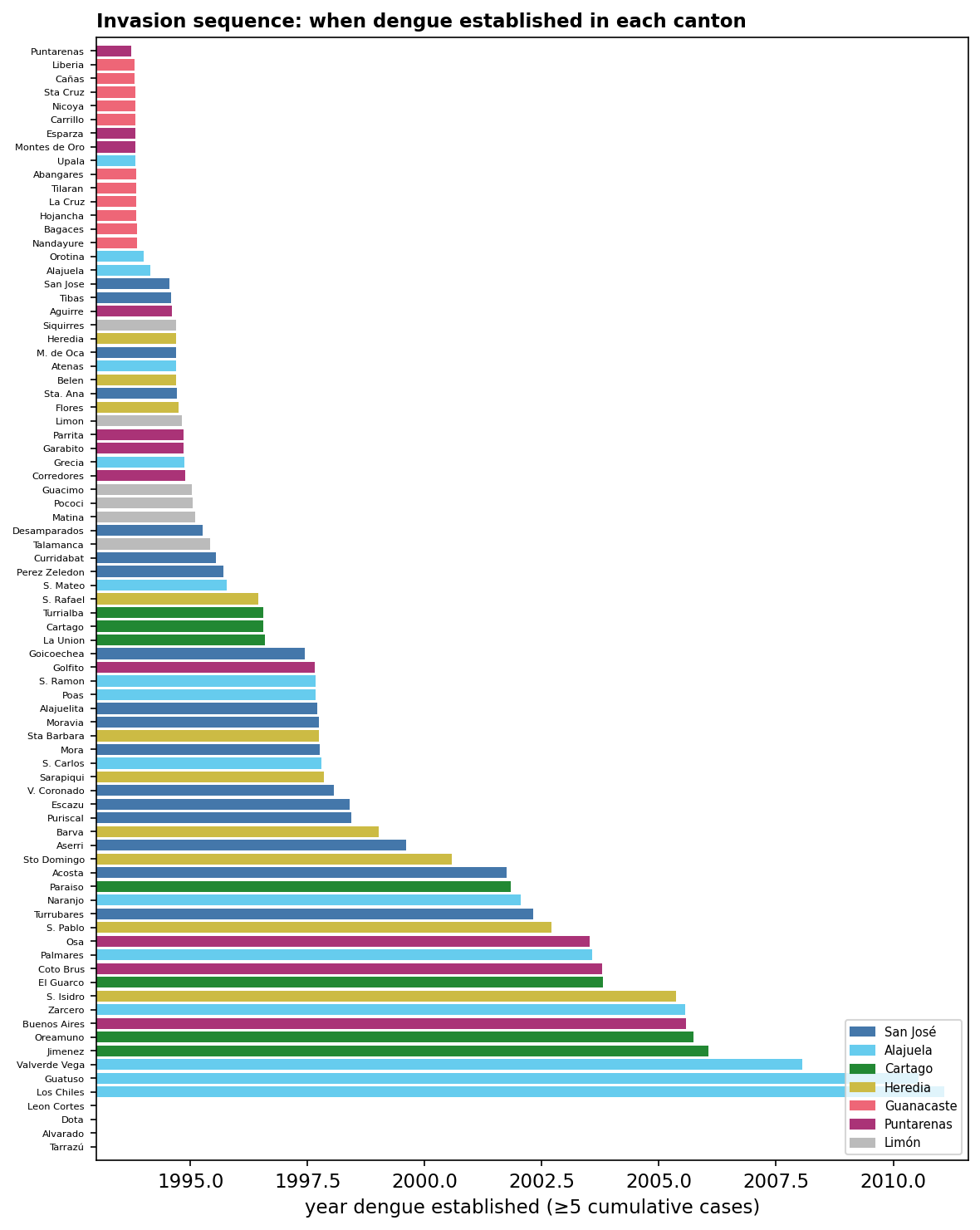}
  \caption{Invasion sequence: the year each canton first accumulated at least five reported cases, colored by province. Lowland Pacific cantons establish first; the highest-elevation cantons (bottom) never reached the threshold.}
  \label{fig:invasion}
\end{figure}

Over the two decades, the national epidemic curve shows the 1993 reintroduction followed by recurrent, multi-year epidemic cycles with major peaks in 1994, 1997, 2003, 2005, 2007, and 2010 (Supplementary Fig.~S1). The full space--time record (Figure~\ref{fig:spacetime}) makes the spatial pattern explicit: dengue established first in the lowland Pacific northwest (Guanacaste and Puntarenas) in 1993--1994, persisted along both coasts through the 1990s, and only intensified in the Central Valley provinces (San Jos\'e, Alajuela, Heredia, Cartago) from about 2003 onward. The order of establishment (Figure~\ref{fig:invasion}) confirms this: coastal Guanacaste and Puntarenas cantons crossed the establishment threshold first, the interior and higher-elevation cantons years later, and the highest, coolest cantons (Tarraz\'u, Dota, Le\'on Cort\'es, Alvarado) never established at all, consistent with the temperature limits of \emph{Aedes} transmission \cite{Mordecai2017}.

\subsection{Seasonality and burden by region}

Dengue is strongly seasonal in every province, peaking in the second half of the year, with the largest mean weekly counts in the North Pacific (Guanacaste) and Caribbean (Lim\'on) regions and the smallest in the cool Central Valley province of Cartago (Supplementary Fig.~S2). Total burden is concentrated in the lowland provinces throughout, but its composition shifts over time as the Central Valley provinces contribute a growing share from the mid-2000s (Supplementary Fig.~S3), the same endemicization visible in the space--time record.

\subsection{The canton-to-canton spread map}

\begin{figure}[H]
  \centering
  \includegraphics[width=0.8\textwidth]{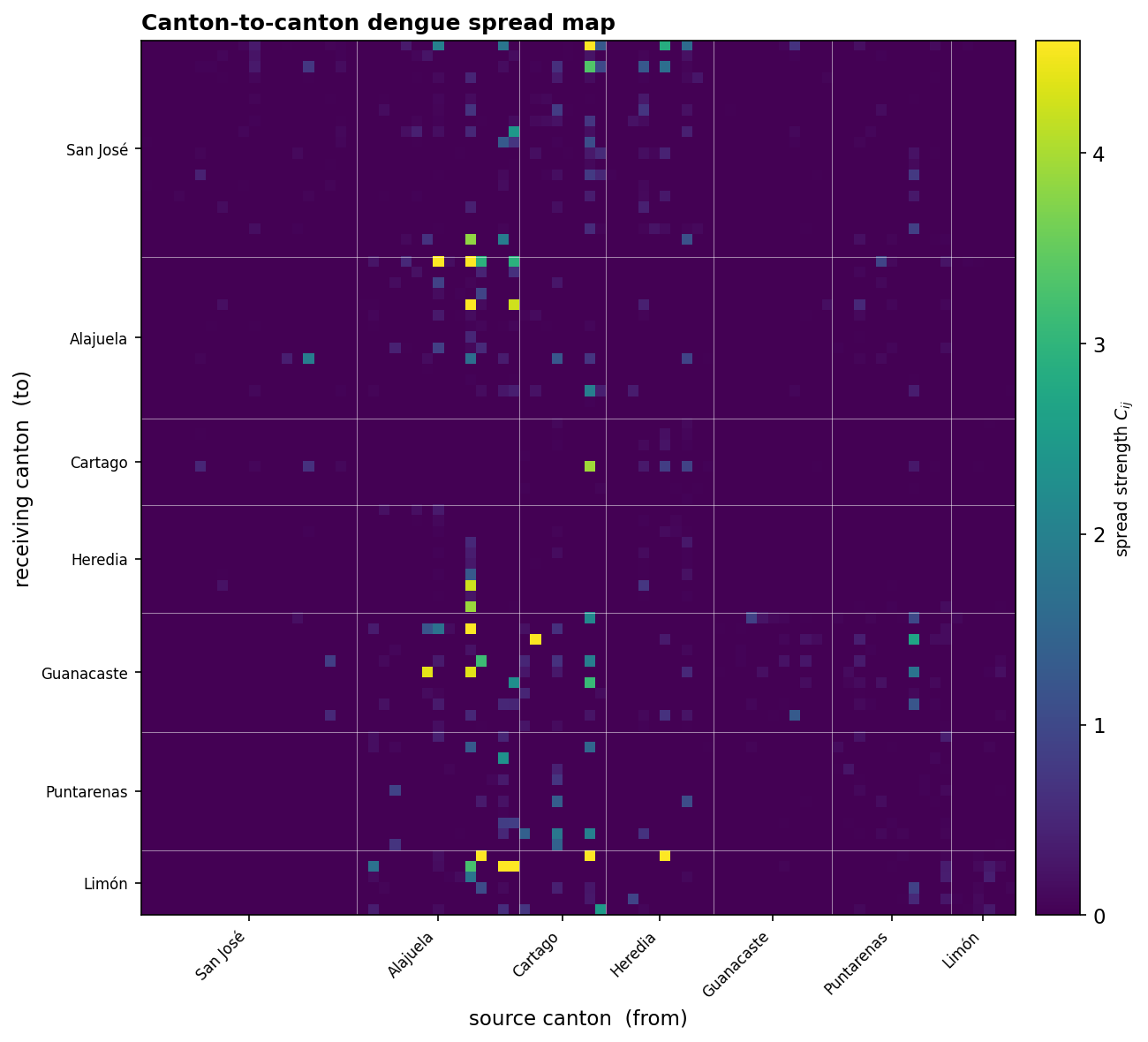}
  \caption{Canton-to-canton dengue spread map $C$ (off-diagonal, grouped by province): the strength with which cases in a source canton (column) are followed by cases in a receiving canton (row). The map is strongly asymmetric; spread has direction.}
  \label{fig:spreadmat}
\end{figure}

\begin{figure}[H]
  \centering
  \includegraphics[width=\textwidth]{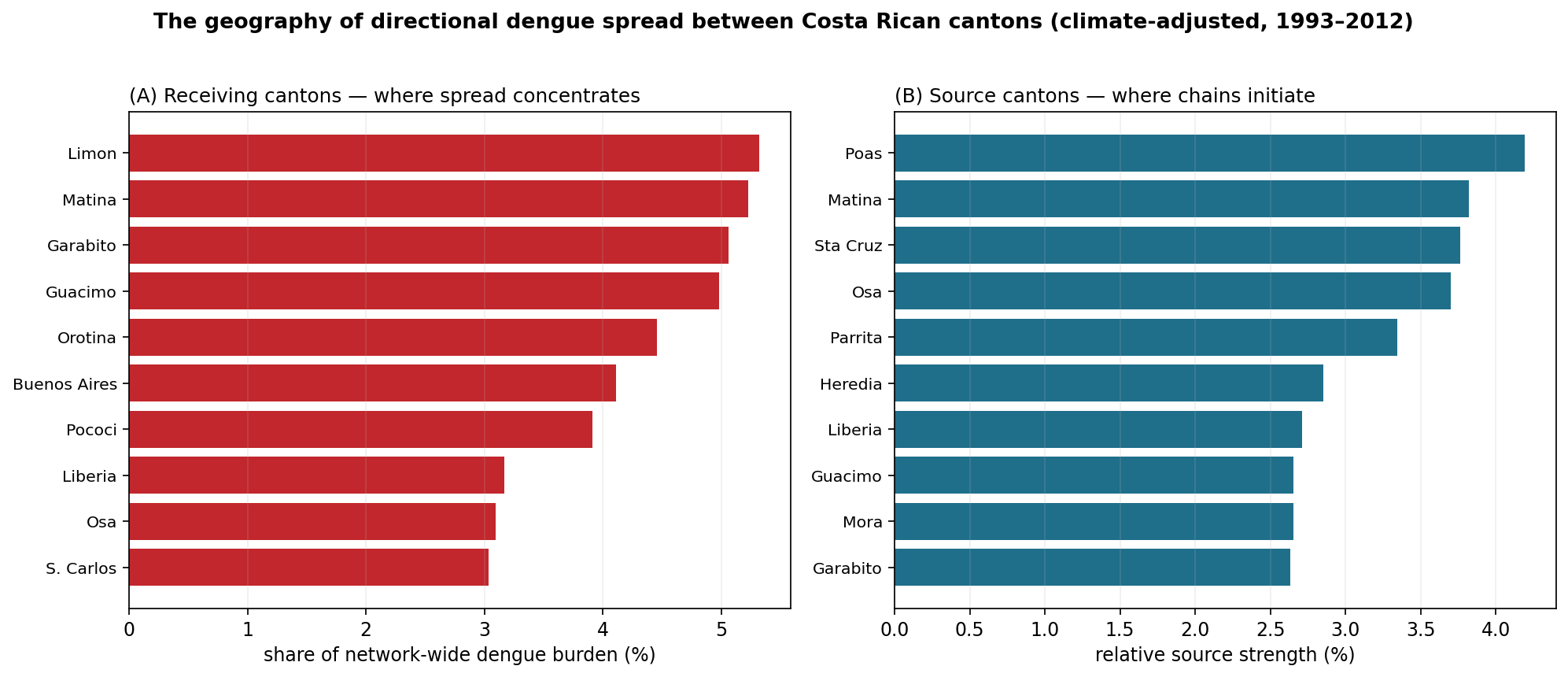}
  \caption{Receiving (A) and source (B) cantons from the climate-adjusted spread operator. Receivers are the lowland Caribbean and Pacific foci and are robust; the source ranking is noisier and should be interpreted with caution.}
  \label{fig:hubsink}
\end{figure}

The reconstructed spread map (Figure~\ref{fig:spreadmat}) is markedly asymmetric: cases in certain cantons are preferentially followed by cases in others rather than reciprocally. Removing the shared-climate signal leaves this directional structure in place, and summarizing the resulting climate-adjusted operator by its dominant pattern, the cantons that most consistently \emph{receive} spread are the hot, low-elevation zones of the Caribbean and central Pacific coasts, foremost Limón, Matina, Gu\'acimo, Garabito, and Orotina (Figure~\ref{fig:hubsink}A), coinciding with the country's established dengue foci and robust to analytical choices. These top receivers were stable across levels of climate-signal removal, though the full canton ranking was only moderately correlated across specifications (Spearman $\rho \approx 0.76$--$0.85$). The identification of \emph{source} cantons (Figure~\ref{fig:hubsink}B) is noisier and should be interpreted with caution.

\subsection{Sources, receivers, and pathways}
\label{sec:sources}

Ranking cantons by the difference between their outgoing and incoming spread strength (Supplementary Fig.~S4) separates net sources from net receivers, and the strongest directed pathways can be drawn as a network (Supplementary Fig.~S5) in which high-burden coastal cantons such as Puntarenas act as prominent hubs. We stress that the receiver side of this picture is the robust one; the source side is more sensitive to reporting in small cantons and should be read as indicative.

\subsection{A local outbreak is amplified across the network}

Although overall transmission over the two decades hovers at or below the replacement level (the system is not, on average, growing), a dengue anomaly introduced into one canton is amplified roughly three- to fourfold as it propagates through the network before subsiding. In practical terms, a local introduction does not stay local: the between-canton coupling in the estimated operator carries and magnifies it beyond what the canton's own transmission would produce, and beyond what a single national reproduction number (which would read as ``at or below threshold'') would lead one to expect. This amplification is modest but robust: it persists at three- to fourfold regardless of how aggressively the shared-climate signal is removed (Supplementary Fig.~S6), and it is the part of the between-canton association that is independent of shared climate.

\subsection{Spread became less efficient as dengue became endemic}

\begin{figure}[H]
  \centering
  \includegraphics[width=\textwidth]{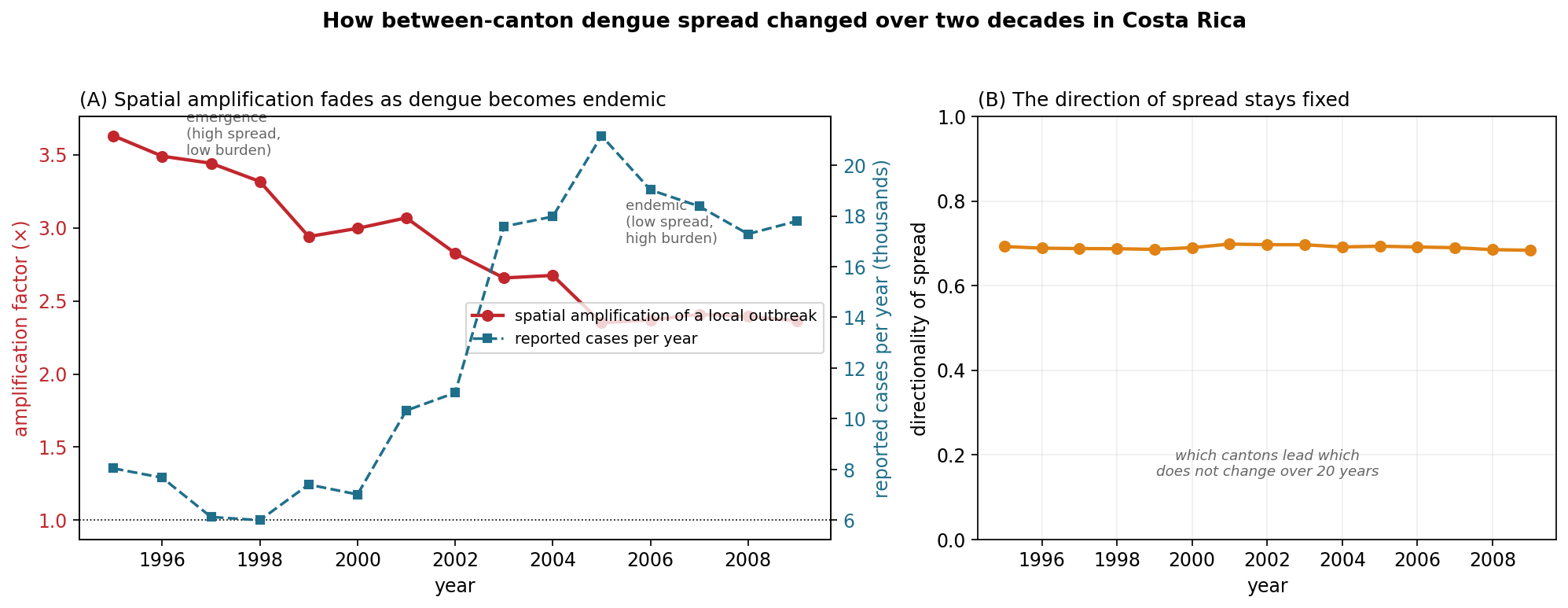}
  \caption{How between-canton spread changed, from the climate-adjusted operator. (A) The amplification factor (red) declined as reported annual cases (blue) rose. (B) The directionality of spread stayed fixed; the pathways did not change, only the strength of amplification.}
  \label{fig:time}
\end{figure}

Tracking the network over five-year windows reveals a clear change (Figure~\ref{fig:time}). The amplification factor declined through the first decade of the record and then plateaued from the mid-2000s, and it did so \emph{as the annual case burden rose}: the network was most efficient at spreading dengue during the low-burden emergence phase of the 1990s, and least efficient once the disease became hyperendemic in the 2000s (Figure~\ref{fig:time}A). The \emph{direction} of spread, by contrast, did not change (Figure~\ref{fig:time}B): which cantons led which remained a stable feature throughout, and only the strength of amplification faded.

\subsection{The decline is robust to changing surveillance}
\label{sec:survcontrol}

\begin{figure}[H]
  \centering
  \includegraphics[width=\textwidth]{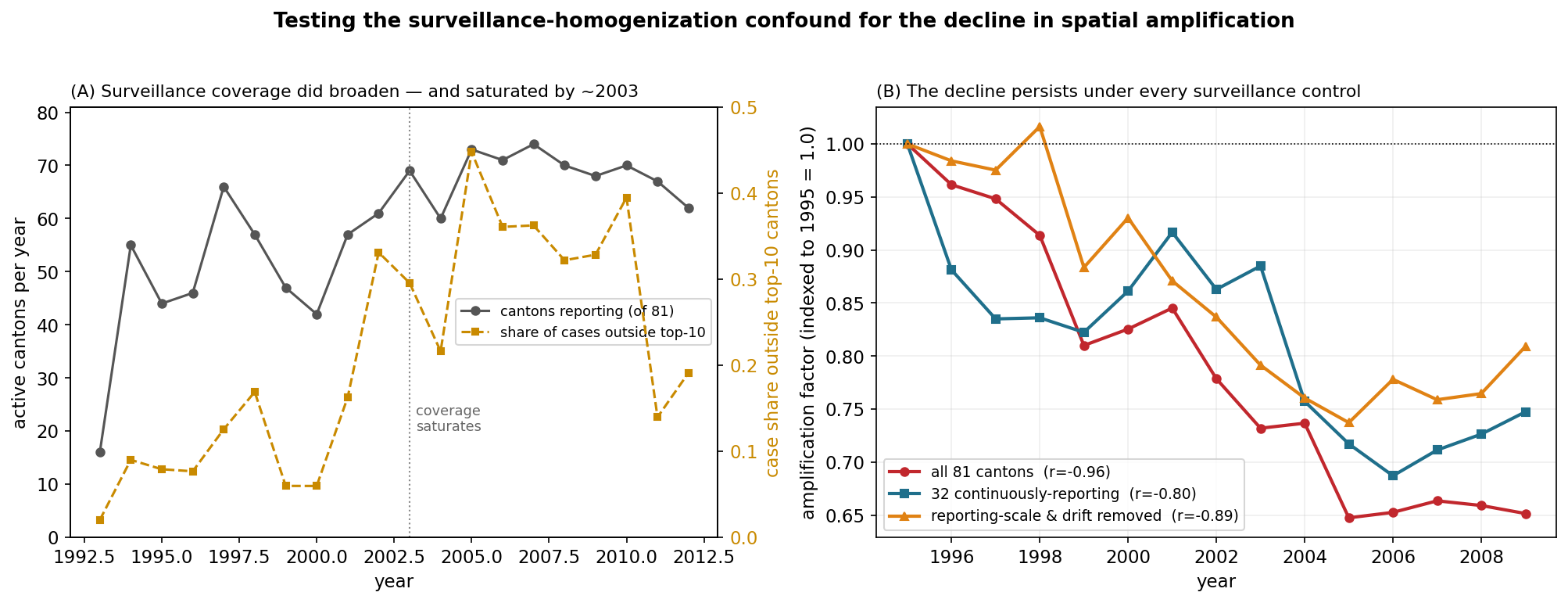}
  \caption{Testing the surveillance-homogenization confound. (A) Reporting coverage broadened over the period but saturated by about 2003. (B) The climate-adjusted amplification factor, indexed to its first window, declines under every control: for all 81 cantons, for the 32 continuously-reporting cantons, and after removing each canton's reporting scale and within-window drift. The relative decline (27--35\%) and its monotonicity are preserved throughout.}
  \label{fig:survcontrol}
\end{figure}

The temporal decline invites an obvious objection: canton-level reporting broadened substantially over the study period, from 16 cantons reporting in 1993 to about 70 by the mid-2000s (Figure~\ref{fig:survcontrol}A), and broadening coverage could, in principle, reduce the apparent between-canton spread for reasons of measurement rather than epidemiology. We tested this directly in three ways (Figure~\ref{fig:survcontrol}B). First, coverage broadened but saturated by about 2003, whereas the amplification factor keeps falling for a further two years before leveling off around 2005; the coverage change and the amplification change are thus not aligned in time, and a one-off shift in coverage cannot by itself produce the observed trajectory. Second, restricting the analysis to the 32 cantons that reported in at least 18 of the 20 years (cantons whose coverage did not change), the decline persists and remains strongly monotonic, with a $31\%$ relative reduction and a correlation of $-0.80$ with time. Third, re-estimating after standardizing each canton's series and removing its within-window trend, which strips out canton-specific reporting scale and slow drift, leaves the decline essentially intact, a $27\%$ reduction with correlation $-0.89$. Across all three specifications, the relative decline is comparable to the $35\%$ seen in the full panel. Expanding surveillance, therefore, accounts for only a modest share of the raw magnitude; a genuine decline in spatial amplification remains even after controlling for it.

\section{Discussion}

Routine dengue surveillance, analyzed with attention to timing and to the confounding effect of shared weather, yields a map of the directional associations by which dengue appears to move between Costa Rican cantons. Three findings are of direct operational relevance. First, between-canton spread is directional and concentrates in identifiable lowland Caribbean and Pacific cantons; these are not only high-burden foci but consistent \emph{receivers} of spread from elsewhere. Second, the estimated canton network amplifies localized outbreaks roughly three- to fourfold even when overall transmission is not growing, so a single introduction may propagate along consistent pathways beyond its point of origin. Third, this amplification was strongest during the emergence phase and has since faded, while the pathways of spread have remained fixed.

These findings fit together as one picture. The sequence in which dengue established across the country, first in the Pacific-northwest and Caribbean lowlands, years later in the interior, and never in the highest and coolest cantons, traces the same lowland geography that the spread map identifies as the persistent receivers, seen through the initial invasion rather than through steady-state spread. One plausible reading of the decline in amplification as the disease matured is a transition between regimes: during establishment, spread may have been dominated by introductions carried from canton to canton, so that the network strongly amplified local outbreaks while total burden stayed modest; once dengue became self-sustaining almost everywhere, transmission would have become locally driven and spatially saturated, and the network's capacity to amplify a fresh introduction would fade even as case numbers rose. This reading is consistent with the spatial invasion record, but it is not the only one, and our data cannot adjudicate among the alternatives. In particular, the two decades span a substantial build-up of population immunity and, very likely, serotype introductions and replacement; these immunological changes reshape transmission in their own right and could produce a declining apparent amplification with no change in spatial coupling at all. Because the surveillance record carries no serotype or immunity information, we cannot separate that pathway from the invasion-to-saturation account, and we therefore offer the latter as a hypothesis suggested by the spatial pattern rather than as an established mechanism. That the \emph{direction} of spread nonetheless remained fixed over two decades points to a durable underlying cause. We conjecture that the pathways reflect the stable structure of human movement (the road network, routine commuting, and seasonal travel toward the coastal lowlands) rather than year-to-year epidemic conditions; because we analyzed no mobility data, this is a hypothesis to be tested rather than a result of the present study. Persistent, movement-structured spatial hierarchies of this kind, in which infection propagates along stable routes from a limited set of driving locations, have been documented for other infections \cite{Grenfell2001}; our results suggest that dengue in Costa Rica follows an analogous, geographically stable template.

Climate is only one of the forces that shape dengue risk. Temperature and rainfall set where conditions permit transmission, but they do not, on their own, determine where an established outbreak will travel next; that also depends on how places are connected. The present analysis addresses this second question directly, recovering from routine surveillance the directional associations along which dengue appears to spread once conditions permit, information that is not contained in local climatic suitability and that we obtained precisely by removing the shared-climate signal. Used in planning, the spread map identifies the receiving cantons and pathways along which risk is realized and propagated, whatever the source of that risk; a canton that is both climatically favorable and a strong receiver in the spread network warrants a different level of attention than one identified by either consideration alone.

\subsection{Implications for surveillance and control}

The findings suggest treating surveillance and pre-emptive vector control as a network problem rather than a set of independent local problems. Enhanced surveillance and early source reduction in the cantons our analysis identifies as receivers, and along the pathways that feed them, might help detect and interrupt spread before it manifests as cases; we emphasize the receiving side deliberately, because our analysis identifies receivers far more robustly than sources (Section~\ref{sec:sources}), so interventions aimed at putative source cantons would rest on weaker evidence; and because amplification was greatest while the disease was establishing, the analysis is consistent with concentrating rapid containment during emergence rather than treating early, low-burden years as low priority. We offer these as planning considerations rather than operational prescriptions, for three reasons. First, the map is historical: it describes how spread was organized over 1993--2012, and although the pathways were stable across that record, using them prospectively assumes that stability persists and should be validated before it is relied upon. Second, the amplification factor is a window-averaged summary of a five-year epoch rather than a real-time indicator of current outbreak risk, so it informs where and when to prioritize rather than triggering specific responses. Third, the analysis rests on a single country's surveillance record, so its transferability to other settings is a hypothesis rather than a demonstrated result. With those caveats, the findings add a spatial and temporal targeting dimension to long-standing recommendations for entomological surveillance-guided integrated vector management and climate-informed early warning \cite{Achee2015}.

\subsection{Limitations}
\label{sec:limits}

Several limitations bound the conclusions. \textbf{Surveillance consistency over time.} Because canton reporting broadened over the study period, we tested directly whether the temporal decline is an artifact of changing surveillance (Section~\ref{sec:survcontrol}, Figure~\ref{fig:survcontrol}). The decline is not eliminated by restricting to continuously reporting cantons, nor by removing per-canton reporting scale and drift, and it continues after coverage saturates around 2003; these controls attenuate its magnitude only modestly. We nonetheless report the decline conservatively: these controls address the broadening \emph{breadth} of reporting but not possible changes in case definitions, diagnostic practice, or reporting thresholds over two decades, and the available data cannot fully separate residual epidemiological change from such measurement change. \textbf{No serotype information (the principal caveat on the temporal decline).} The record does not distinguish dengue serotypes, so serotype introductions and replacement, and the build-up of serotype-specific and cross-reactive immunity, cannot be examined here. These are plausible drivers not only of the amplification we measure but of its decline: immunological change over the two decades could by itself produce a falling apparent amplification independent of any change in spatial coupling. Molecular studies of Costa Rican outbreaks have characterized the serotypes circulating during major epidemics \cite{SotoGarita2016}, but a serotype-resolved and immunity-resolved incidence series would be needed to distinguish this pathway from the spatial regime transition we hypothesize. \textbf{Aggregate, ecological data.} The analysis operates on reported case counts at the canton level and describes population-level spread, not individual transmission events; source-canton identification in particular is sensitive to reporting in small, low-incidence cantons and is reported cautiously. \textbf{Climate adjustment is indirect.} We removed shared climate through its statistical signature rather than through measured variables; a stronger analysis would use high-resolution temperature and precipitation fields as explicit adjustments, leaving spread net of measured climate. That refinement is the natural next step and would sharpen the separation between directional between-canton spread and residual climatic co-movement. \textbf{Associational, not mechanistic.} The analysis is a lagged regression on reported case counts, so the directional associations it recovers are consistent with connectivity-driven spread but do not on their own establish it. Factors correlated with weather yet not fully removed by subtracting the seasonal cycle and the leading modes of co-movement, such as human mobility, spatially structured changes in healthcare-seeking or reporting, or coordinated vector-control activity, could contribute to the residual associations. We therefore frame the findings as directional associations that remain after accounting for shared climate, and treat the connectivity interpretation as a hypothesis to be tested directly with mobility and intervention data.

\section{Conclusions}

Two decades of routine dengue surveillance show that between-canton spread in dengue incidence is directional, concentrated in specific lowland foci, and that the estimated network can amplify a local outbreak several-fold, and that this amplifying capacity was greatest when the disease was emerging and has since faded, even as burden rose, while the pathways of spread stayed fixed. These historical patterns offer a basis for prioritizing surveillance and pre-emptive vector control toward the cantons and pathways that most often received dengue, and for weighting the response toward the early, high-spread phase of a new introduction, provided such prospective use is validated rather than assumed. Because it is built entirely from routine case reports, the approach could be applied wherever dengue is reported at a consistent administrative unit, though its transportability remains to be demonstrated.

\section*{Acknowledgements}

The author thanks the Centro de Investigación en Matemática Pura y Aplicada (CIMPA), the Escuela de Matemática of the Universidad de Costa Rica, and the Costa Rican Ministry of Health for compiling and sharing the dengue surveillance data.

\newpage
\section{Supplementary material}
\setcounter{figure}{0}
\renewcommand{\thefigure}{S\arabic{figure}}
Supplementary Figures~\ref{fig:S1}--\ref{fig:S6} accompany this article.
\vspace{1em}
 
\begin{figure}[H]
  \centering
  \includegraphics[width=\textwidth]{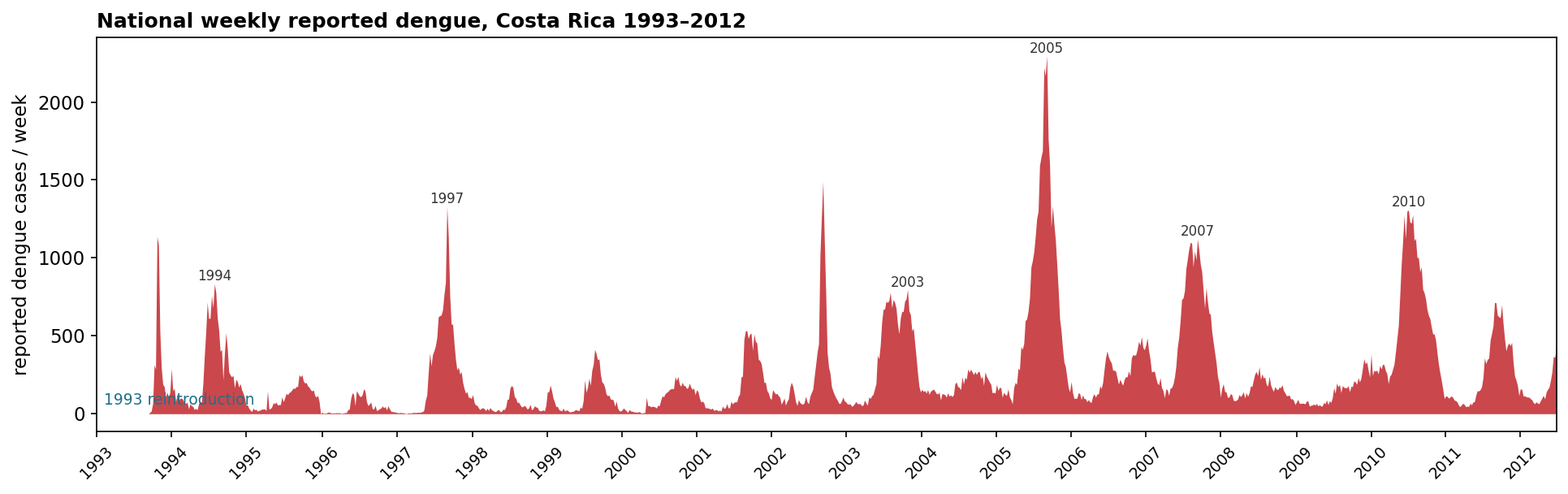}
  \caption{\textbf{National weekly reported dengue cases}, Costa Rica 1993--2012, following the 1993 reintroduction. Labels mark the major epidemic years.}
  \label{fig:S1}
\end{figure}
 
\begin{figure}[H]
  \centering
  \includegraphics[width=\textwidth]{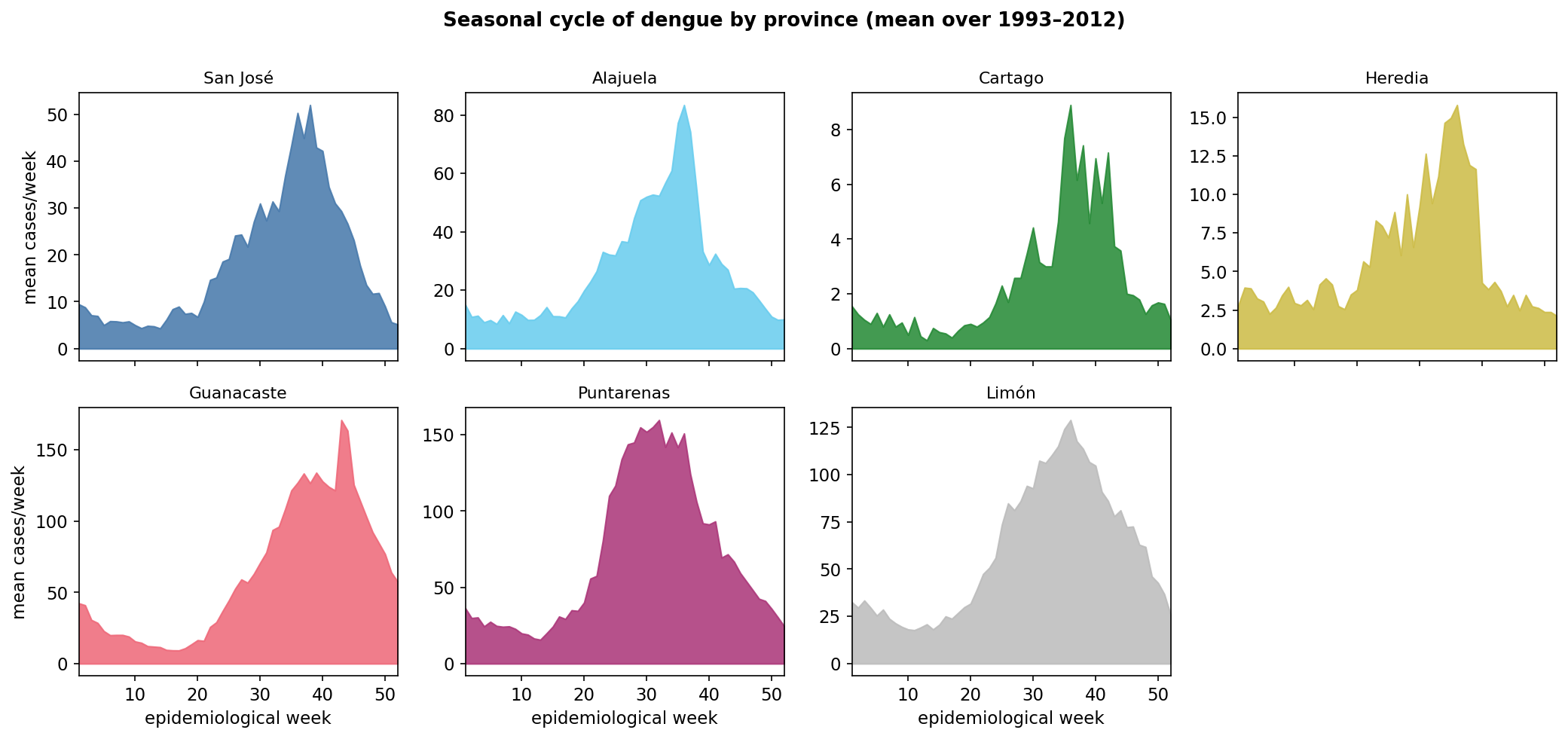}
  \caption{\textbf{Seasonal cycle of dengue by province} (mean weekly cases over 1993--2012). Note the differing vertical scales; all provinces peak in the second half of the year.}
  \label{fig:S2}
\end{figure}
 
\begin{figure}[H]
  \centering
  \includegraphics[width=\textwidth]{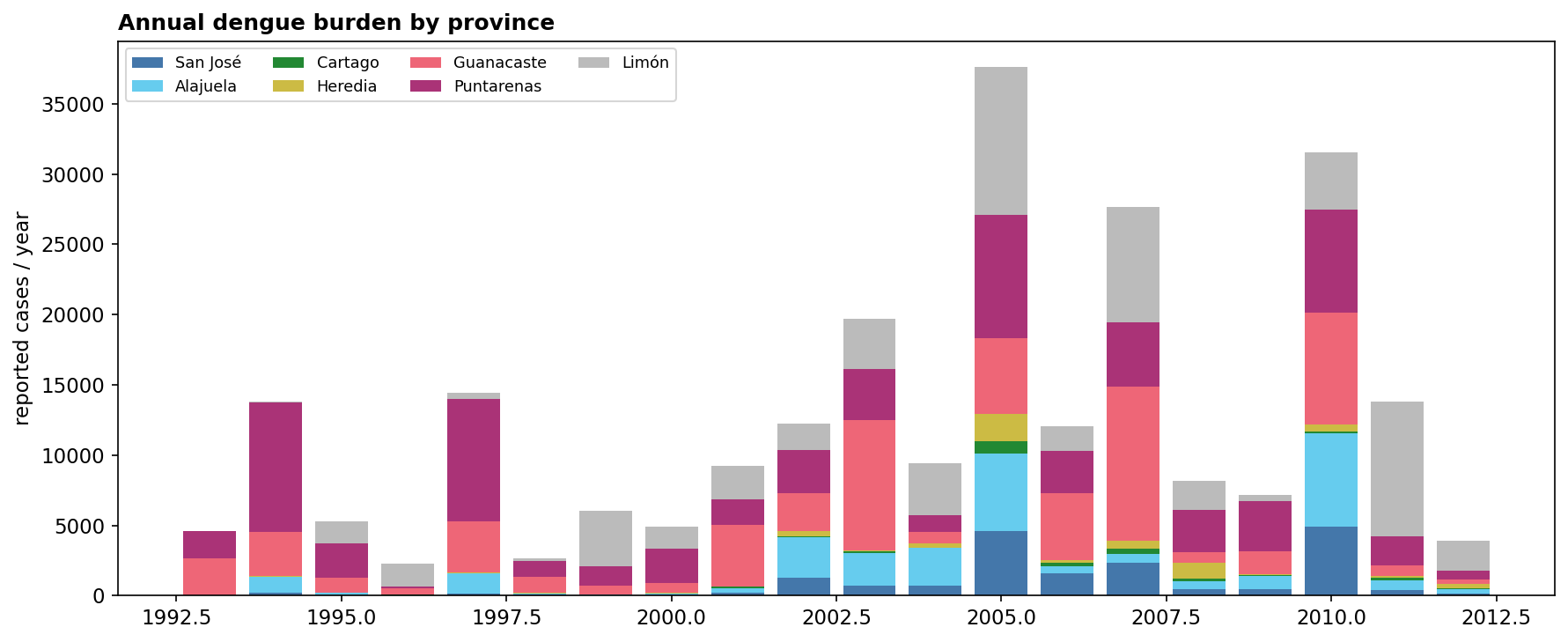}
  \caption{\textbf{Annual reported dengue by province.} Burden is dominated by the lowland provinces, with a growing Central Valley contribution as the disease became endemic.}
  \label{fig:S3}
\end{figure}
 
\begin{figure}[H]
  \centering
  \includegraphics[width=0.78\textwidth]{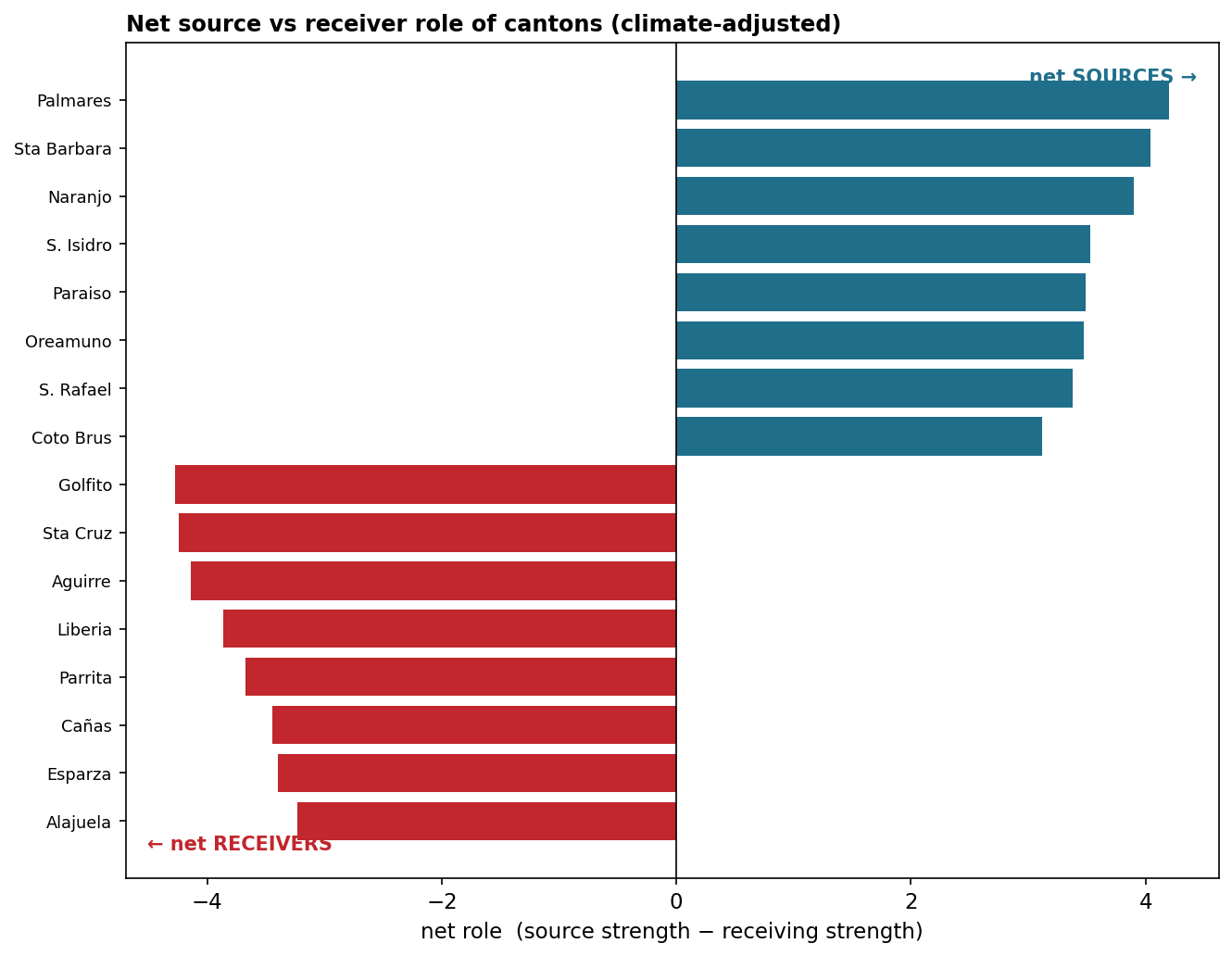}
  \caption{\textbf{Net role of cantons:} outgoing minus incoming spread strength in the climate-adjusted map. Blue cantons act as net sources, red as net receivers. The source side is sensitive to reporting from small cantons and is interpreted with caution.}
  \label{fig:S4}
\end{figure}
 
\begin{figure}[H]
  \centering
  \includegraphics[width=\textwidth]{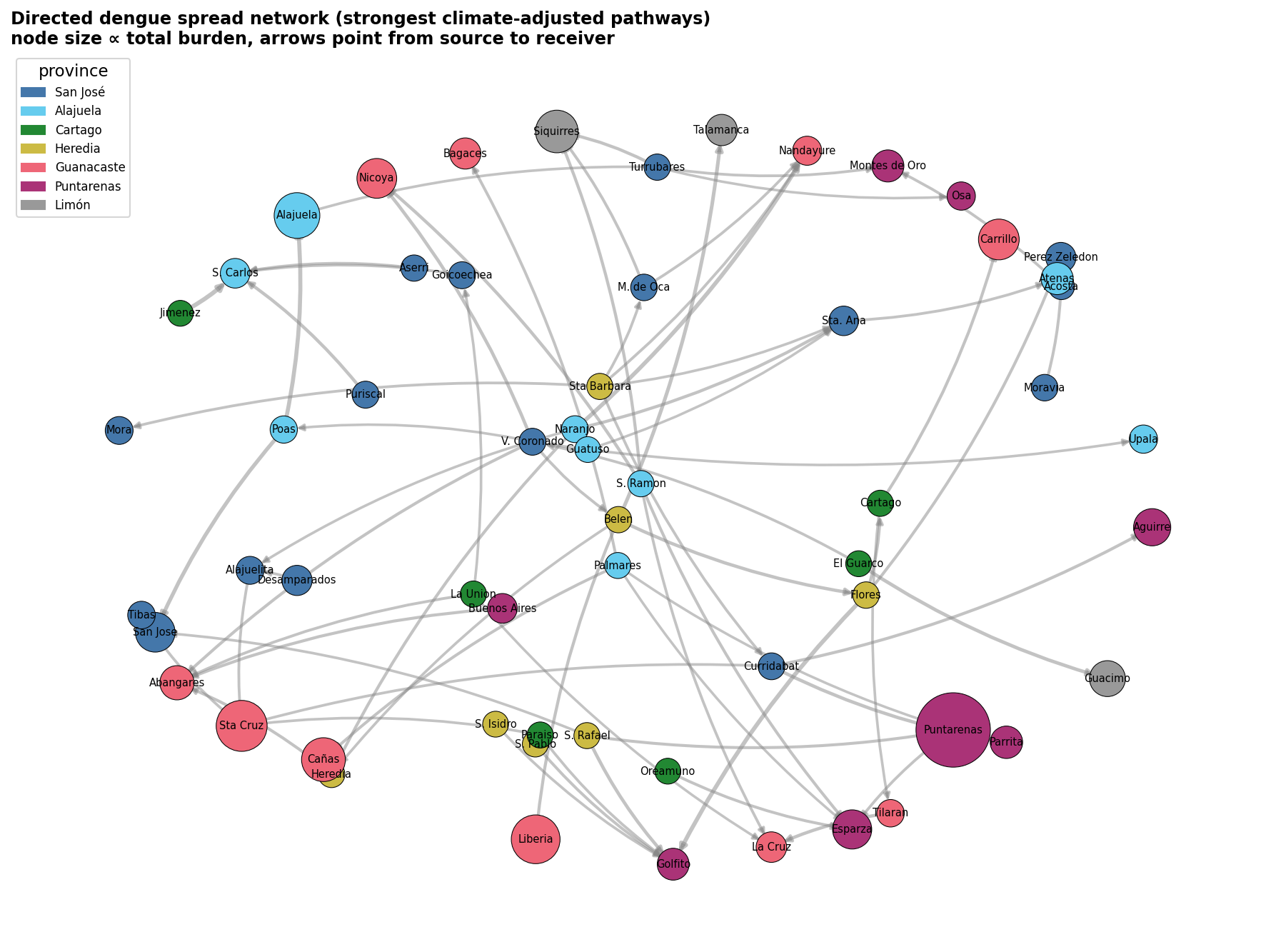}
  \caption{\textbf{Directed dengue spread network:} the strongest climate-adjusted pathways between cantons. Node size is proportional to total dengue burden, node color denotes province, and arrows point from source to receiver. This figure is illustrative; the source side of the network is the less robust part of the analysis.}
  \label{fig:S5}
\end{figure}
 
\begin{figure}[H]
  \centering
  \includegraphics[width=0.72\textwidth]{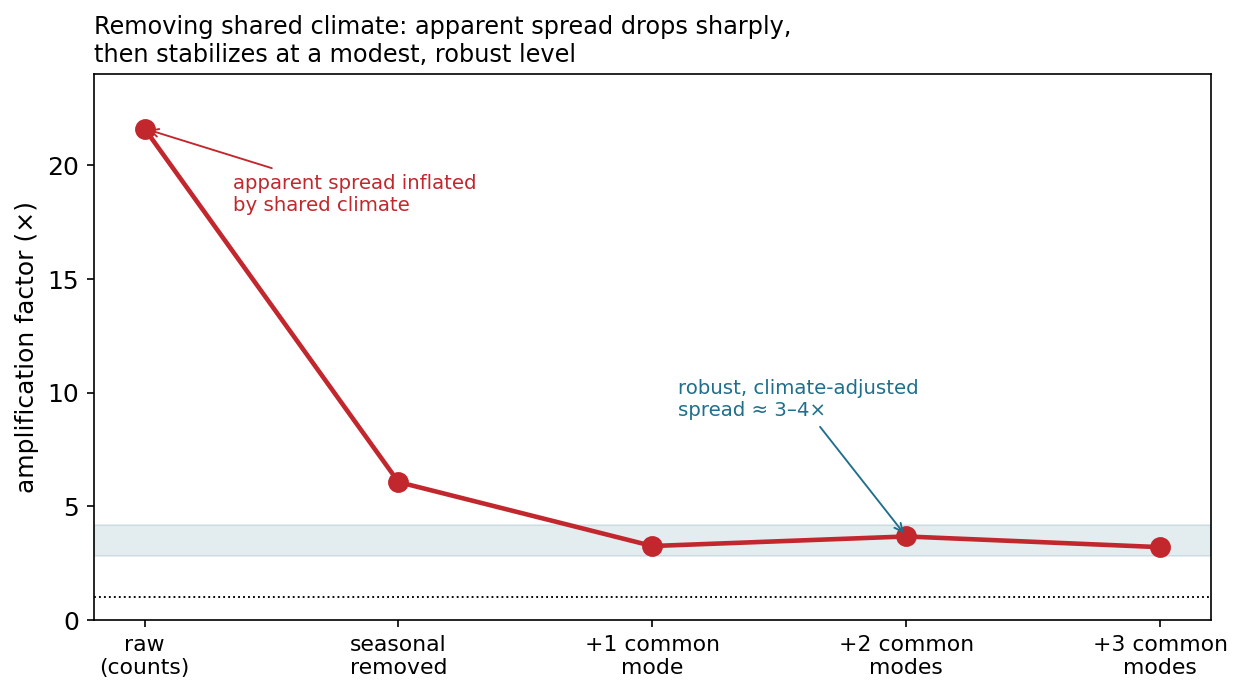}
  \caption{\textbf{The directional spread is robust to climate adjustment.} As the shared seasonal and climatic signal is progressively removed, the apparent strength of spread drops sharply from its raw (climate-confounded) value and then stabilizes at a modest, direction-specific amplification of about three- to fourfold, regardless of how much common signal is removed.}
  \label{fig:S6}
\end{figure}

\end{document}